# Directed Droplet Motion — Its Versatile Nature and Anticipated Applications

Panagiotis E. Theodorakis[a,1], Andrey Milchev[b,2]

[a]Institute of Physics, Polish Academy of Sciences, Al. Lotników 32/46, 02-668 Warsaw, Poland

[b]Institute of Physical Chemistry, Bulgarian Academy of Sciences, G. Bonchev Str. Bl. 11, 1113 Sofia, Bulgaria

Abstract

Applications such as digital microfluidics and bio-diagnostics rely on droplet locomotion. A prominent example of such motion is durotaxis, a phenomenon that requires a stiffness gradient along a surface for the transport of liquids, cells, or other nano-objects. Using surfaces with varying properties in specific directions can be exploited as a universal concept for fluid transport with or without external energy supply. Changes in properties may refer to substrate patterns, Laplace pressure changes, wettability gradients, etc., leading to exciting phenomena, which can be employed in novel applications in various technologies. Here, we report on key results and progress in the area of directed droplet motion over the years, and we provide perspectives and implications for anticipated applications.

Contents

[1] panos@ifpan.edu.pl

[2] milchev@ipc.bas.bg

1. Introduction

A number of phenomena in nature relate to the spontaneous motion of liquid droplets on various types of substrates, for example, substrates with different patterns, wettability, roughness, *etc.* [1–4]. By observing nature at work [3], we gain inspiration that can be transformed into innovations and sustainable technologies. More specifically, causing and controlling fluid (droplet) motion can be used in applications, such as printing, micro-reactors, cargo transport, bioanalysis, electricity generation, heat transfer, water harvesting, desalination, self-cleaning, and antifogging [5].

Despite fundamental differences among such a wide spectrum of applications, various common characteristics can be identified. For example, droplets are often smaller in size than the capillary length $\ell_c = \sqrt{\frac{\gamma_{lv}}{\rho g}}$, reflecting the length scale when gravity effects start to take over as the size (mass) of the droplet increases. Here, $\gamma_{lv}$ denotes surface tension, $\rho$, density, and $g$ is the acceleration due to gravity. The interfacial interactions between the fluid and the air or other medium (*e.g.*, the substrate or walls of a microfluidic capillary) are expected to play a key role at small scales (smaller than $\ell_c$) and determine the outcome of the motion. An illustrative example is presented in Figure 1 assuming a chemically inhomogeneous surface [6]. For this and cases with thermal gradients, which includes the Marangoni contribution, details can be found in Ref. [6]. When the drop size $l < \ell_c$, the radius of curvature is constant and the drop adopts a spherical-cap shape. In this case, uncompensated dynamic Young forces at the leading and trailing edges of the droplet will drive the droplet towards larger spreading coefficient, $S = \gamma_{sv} - \gamma_{sl} - \gamma_{lv} < 0$ values, where $\gamma_{sv}$, $\gamma_{sl}$, and $\gamma_{lv}$ are the solid–vapour, solid–liquid, and liquid–vapour interfacial tensions, respectively. In contrast, when $l > \ell_c$, the drop cap is flattened due to gravity and resembles a pancake structure. In this case, $S$ increases in the $X$ direction and the top surface has a small slope, creating a pressure gradient that forces the liquid to flow (Figure 1). The above two scenarios would result in the velocity profiles shown in Figure 1 and constitute fundamental mechanisms for droplet motion on substrates with varying properties (e.g., wettability gradients, topographical patterns, *etc.*).

The interactions at the interface between the liquid and the solid are crucial for the droplet motion. These can fundamentally be explained at a molecular level, but, at the same time, they determine the efficiency of practical applications beyond the molecular scales. By designing and modulating the properties of a substrate (*e.g.*, its topography) for various fluids (*e.g.* surfactant-laden droplets), the necessary properties of the system can further be fine-tuned for the specific application in mind. In view of the large number of possible modifications and designs that can lead to the aforementioned force imbalance, a variety of systems can be conceived and realized experimentally. For this reason, the scope of research in the area of directed motion is also broad, including biological systems (*e.g.*, cells moving onto tissues) and those that mimic them (biomimetic interfaces) [7]. In the context of biological systems, for example, the cell migration is governed by the extracellular matrix stiffness and can take place in the same (*durotaxis*) or the opposite (*anti-* or *negative durotaxis*) direction of stiffness increase along the substrate (Figure 2). Here, however, the cell movement may also rely on mechanosensors, such as YAP/TAZ, nucleus and piezo1/2, thus involving active processes [7].

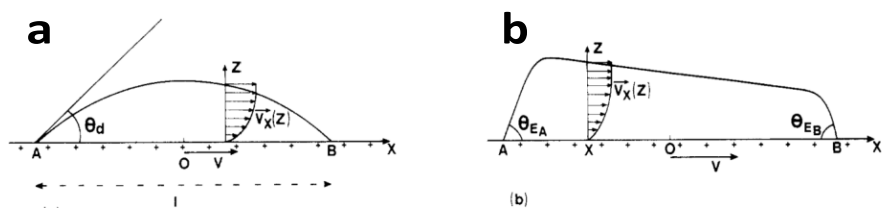

Figure 1: Drop on a chemically inhomogeneous surface. (a) If $l < \ell_c$, the radius of curvature is constant and the droplet has a circular cross-section. In this case, the dynamic angle, $\theta_d$, is smaller than the equilibrium contact angle at A and larger than the equilibrium contact angle at B during motion in the $X$ direction. Uncompensated dynamic Young forces drive the droplet toward larger spreading coefficient $S = \gamma_{sv} - \gamma_{sl} - \gamma_{lv}$ < 0 values. The resulting velocity profile, $V_X(Z)$ is shown. (b) A large drop $l > \ell_c$ is flattened by gravity. $S$ increases in the $X$ direction, the top surface has a small slope, which creates pressure gradient forcing the liquid to flow. Reprinted (adapted) with permission from Ref. [6]. Copyright 1989 American Chemical Society.

Although our discussion has been inspired by the durotaxis phenomenon, a term coined for cells moving on tissues by virtue of a stiffness gradient along the substrate [8], here, we will touch upon phenomena dealing with directed droplet motion as a common theme, that is, surfaces with properties that induce a gradual or more abrupt variation along the direction of motion (*e.g. stiffness gradient* in the case of durotaxis). Here, again, the concept of a property variation or a gradient in a specific direction can be broader and include cases where the droplet simply senses a difference in its immediate environment, which suffices to cause the droplet motion. A few examples, relevant for our discussion, are therefore systems with wettability and thermal gradients, Laplace pressure differences, or simply substrates with different physical patterns. One way of looking at this to make a distinction between the various cases is whether causing and sustaining droplet transport requires a continuous supply of energy from an external source (e.g., an external field, mechanical vibration, etc.). Then, one may distinguish two broad categories, namely those of *active* (continuous energy supply or stimulus from an external source is needed) and *passive* (without external energy supply) transport processes. Still, various cases may involve both active and passive scenarios. In addition, although there currently exists a large number of methods to manufacture gradient surfaces [9], the focus of this review will remain on the driving processes and system properties and only a few recent methods for creating surfaces with varying properties will be mentioned. In the context of passive processes, recent studies by the authors will also be discussed along with lessons learned on the mechanisms of durotaxis and more generally directed droplet motion onto both stiff [10–12] and soft substrates [13–15].

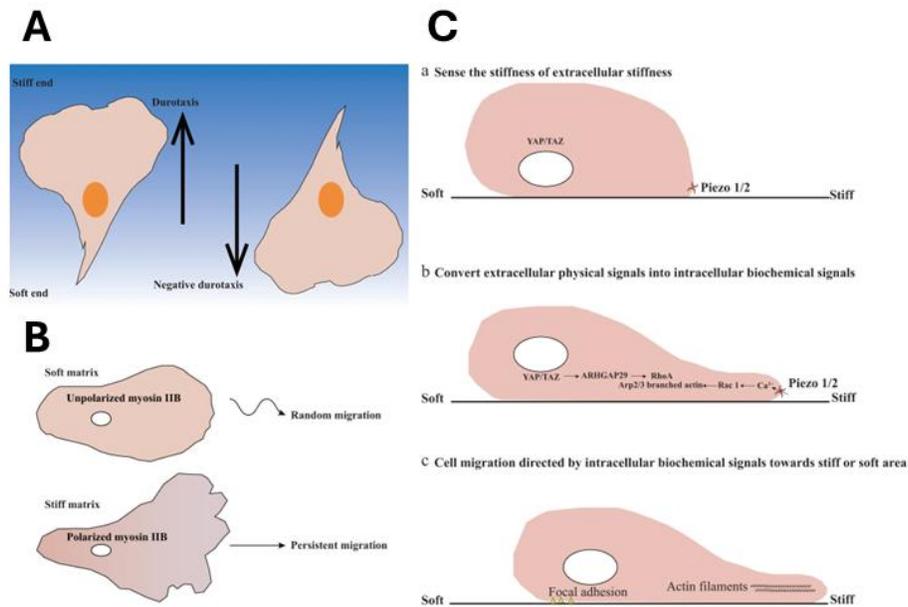

Figure 2: (A) Cell migration is directed by extracellular matrix stiffness. Cell migration from the soft to the stiff end is known as durotaxis, while the migration of cells towards the soft end of the extracellular matrix is called negative durotaxis; (B) Cells with unpolarised myosin IIB tend to exhibit random migration, whereas the ones with polarised myosin IIB show persistent migration. (C) Cells on stiffness gradient gels sense the stiffness of the extracellular matrix through mechanosensors, such as transcriptional regulators YAP/TAZ, nucleus, and piezo1/2. Figures from Ref. [7].

In the following and before delving into the details of directed fluid motion in non-biological contexts in Section 3, we will briefly discuss droplet propulsion in biology in Section 2. Then, Section 3 includes a critical discussion on directed motion by electrical/magnetic, optical, and mechanical actuation, Laplace pressure, Marangoni effects, wettability gradients, as well as motion onto slippery liquid-infused substrates and by active droplets. We will also explore anticipated applications and provide our perspectives in Section 4. We will complete this review by briefly drawing a few broader conclusions in Section 5.

2. Droplet propulsion in biology

One fascinating process in which cells migrate along gradients of stiffness of the extracellular matrix is *durotaxis* [8, 16]. It plays a constructive role in various biological processes, such as embryonic development, tissue homeostasis, and regeneration [9, 17], but it can also lead to pathological conditions, such as cancer invasion and fibrosis, where altered tissue stiffness correlates with disease progression [18–21].

While durotaxis commonly refers to motion caused by a stiffness gradient on substrates, it involves complex *mechanosensing* processes in a biological context, which are employed by cells to detect and respond to substrate stiffness changes and have routes to natural processes in an organism [21, 22]. Cells typically experience larger traction forces on stiffer regions of a substrate [23] through interactions that involve the cytoskeleton, focal adhesions, and associated signalling pathways, ultimately leading to maximal traction forces [24]. Although a complete mechanistic description is beyond the scope of this review, it is generally accepted that both *mechanical and chemical processes* [25–30] govern cellular motion. Advances in imaging techniques (e.g. *Brillouin microscopy* [31]) offer promising avenues for probing stiffness gradients and cell migration dynamics in complex biological environments [17], potentially enabling controlled manipulation of cell trajectories via micro-scale stiffness or topographic patterns
[32].

Although durotaxis usually describes the motion from softer to stiffer regions [8], the opposite phenomenon (*negative durotaxis* or *antidurotaxis*) has also been observed (Figure 2) [7, 14, 15, 33], for example, in certain neuron and cancer cells [7, 20], where the migration takes place towards softer tissues. Antidurotactic motion also arises from imbalances in traction forces in the presence of stiffness gradients [23], but its underlying mechanisms remain under debate [7, 19, 34]. Theoretical approaches, such as stochastic clutch models and active gel theory [35], have been proposed to rationalize these behaviours. Moreover, several models suggest that cells generate stronger traction forces on stiffer substrates [23], while others describe durotaxis as an *elastic stability phenomenon* in which cytoskeletal elasticity governs the migration [36].

Inspired by biological durotaxis and adopting a simplistic view, non-biological analogues have been developed for *droplets migration along stiffness gradients*. For example, molecular dynamics (MD) simulations reveal that the propulsion velocity of drops depends on parameters such as droplet size, viscosity, wettability of the substrate and gradient strength [10], where the latter can be precisely engineered [37]. Smaller droplets with lower viscosity typically move faster on more wettable, stiffer regions [10]. These systems demonstrate the feasibility of *gradient-driven motion* at the nanoscale without the expense of *external energy supply* or complex mechanosensory processes, driven by the imbalance of Young forces as a driving force of the motion as discussed in Ref. [6] (Figure 1). In addition to stiffness gradients, *chemotactic interactions* can also drive droplet translocation in active processes. For example, enzymatically *active droplets* can generate local chemical (pH) gradients that can lead to collective, self-organized motion, reminiscent of biological behaviour [38]. Beyond gradient-driven self-propelled motion, droplets can also be steered by the imbalance in *interfacial stresses* due to changes in *surface topology*. The combination of microstructure grooves and surfactants, for example, can cause long-range droplet self-propulsion including the *Marangoni contribution* [6], achieving droplet motion over macroscopic distances [39]. Such strategies provide versatile ways of liquid transport thus transforming biological inspiration into *engineered functional systems* with droplet motion into a certain direction as a common theme.

3. Directed Motion in non-biological context

*Directed motion in non-biological systems* refers to the controlled locomotion of droplets, particles, or other small objects (e.g. nanoflakes, nanosheets, *etc.*) along a path onto a surface. In fluids, and particularly in the case of droplets, this directed behaviour contrasts with random diffusive motion and typically arises from force imbalances along the direction of gradient or varying substrate properties. Such gradients (mechanical, chemical, thermal, electrical, or magnetic) can be externally imposed or self-generated by the droplet, for example, *via* interfacial effects (e.g., tribological) or changes in concentration (e.g., surfactant).

Gradients of surface properties are pivotal in *materials science, nanoengineering* and *robotics*, where precise control of motion (position, velocity, acceleration, and stability) is crucial. Reliable and predictable droplet steering requires well-defined parameters governing the direction, speed, and stability of motion, rendering gradient engineering an essential consideration for application design. In *microfluidic systems*, fluids move through microchannels under applied pressure, electrokinetic forces, or other stimuli to achieve precise control of chemical reactions or cell sorting.[40]. At the nanoscale, *nanorobots* depend on exquisite motion control to perform tasks, such as targeted drug delivery or molecular assembly [41]. These are often guided by *magnetic, acoustic* or *optical gradients*, which may enable navigation with high precision. In more complex systems, such as *nanoparticle-laden fluids*, self-propulsion mechanisms can rely on intrinsic asymmetries in particle design, for example, *Janus particles*, with chemically or physically distinct hemispheres. These can exhibit a directed motion that can be tuned by external fields or local concentration gradients [42]. Similarly, substrates with *programmed stiffness or chemical gradients* can induce droplet locomotion, enabling systems to self-adapt to environmental changes in an adjustable, 'self-programmable' manner.

A variety of *driving forces* can actuate such motion, including electrical and magnetic fields, pressure differences, chemical gradients, and optical forces. These can operate in both *active* (requiring supply of energy) and *passive* (without external energy supply) modes. Still, the distinction between active and passive locomotion is often blurred, since many systems may combine both scenarios or, moreover, may function without explicit gradients, as, for example, in various *ratchet-like motions*, which can be actuated by substrate vibrations [43–46].

The broader perspective pertains to *droplet actuation* in the case of active transport. Actuation techniques may include *electrowetting* (electric fields), *ferrofluid manipulation* (magnetic fields), *mechanical vibration* or pressure actuation, and *thermal* or *optical gradients*. These possibilities underpin technologies, such as *digital microfluidics*, *lab-on-a-chip systems*, and *micro-reactors*, where precise droplet control is essential [47]. In view of the diversity of actuation methods, attention has been paid to *magnetic*, *optical*, and *mechanical actuation*, which have demonstrated significant potential for emerging applications. However, phenomena that involve more complex processes, such as *projectile motion* [48], curvature-driven transport combined with droplet impact [49] *droplet detachment* on bioinspired superhydrophobic surfaces [50, 51] and *vibration-induced contaminant removal* [52] are beyond the scope of this review.

*3.1. Electrical/magnetic actuation*

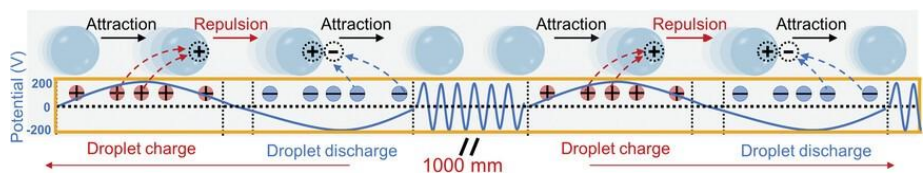

Figure 3: Droplet motion propelled by alternate potentials. Sketch of a droplet charged and discharged in subsequent steps on the electret film with alternate positive and negative potential. Reproduced with permission from Ref. [53]. Copyright 2024 Wiley-VCH GmbH.

Electrical forces can be used to cause and steer the motion of droplets. In particular, electrowetting on dielectrics is a common technique that applies electric fields, which in turn affect the wettability of the substrate [54]. In this case, the variations in wettability along the substrate are exploited to steer the droplet motion. To achieve this, a chemical molecular layer that can generate and store triboelectric charges can be used in the case of triboelectric wetting [55]. Triboelectric charges are generated by friction and stored along the molecular layer, in a way that creates a gradient that can sustain a continuous droplet motion. The output signal of triboelectric nanogenerators can be used to adjust the electric fields to manipulate both macro and micro droplets [56].

Such changes of electric properties of the substrate can be illustrated when a symmetrical waved alternating potential (WAP) on a superhydrophobic surface is applied (Figure 3) [53]. The droplet is propelled by alternating positive and negative potentials, which leads to droplet charging and discharging in subsequent steps with a mechanism that is schematically illustrated in Figure 3. In this case, the propelling forces are either repulsive or attractive, depending on the sign of the potential. This motion has been shown to depend on the charged diameter and the net potential gradient, but the setup provides the possibility of self-propulsion of droplets without distance limits at 'high' velocities, rendering it suitable for liquid diodes and logic gates in microfluidics applications. Varying the charge density and creating charge gradients along specific directions has been explored by MD simulation [57], while electrowetting theory [58] has attempted to explain such phenomena highlighting a positive correlation between charge gradient and the driving force of the droplet motion.

Pyroelectricity and piezoelectricity can act as external stimuli for initiating a taxis-like motion offering versatile ways of droplet manipulation. A specific example here is motion in response to external stimulus by a human finger (piezoelectricity) [59], similar to an electrostatic tweezer [60], which can be used to remotely trap and guide droplets. The Coulomb attraction between the tweezer (or other external stimulus) and the droplet offers precise control on the motion, namely, distance, velocity, and direction in diverse environments, including droplets in oil and on tilted substrates.

In contrast to electric actuation, magnetic actuation requires that the liquid has magnetic properties, which can respond to magnetic fields, for example, ferrofluids. Applying a varying magnetic field on a ferrofluid droplet can lead to its transport, but a pearling phenomenon (droplet breakup) occurs during the motion [61]. However, bidirectional droplet motion has been achieved by using a magnetically actuated superhydrophobic *ratchet* surface (more on ratchets in the context of mechanical actuation below) [62]. In this system, the surface consisted of strips that can be tilted in either direction by means of an external magnetic field and thus steer the droplet motion in different directions.

In the context of nanoscale devices, various studies have proposed different mechanisms, such as nanomotors [63] and those based on a 'flow' of electrons [64]. This concept has been especially applied in carbon nanotubes. In particular, controllable and reversible atomic-scale mass transport of indium along carbon nanotubes has been shown as an example of electromigration [65]. In an attempt to explain this phenomenon, research has suggested that thermal effects (thermomigration) and thermal evaporation also play a significant role with respect to the electromigration force [66]. While these concepts have been inspired by mechanical systems, a different approach has been explored for a four-wheeled molecule on a metal surface, where directional motion was achieved by a series of conformational changes provoked by electrical pulses [67]. To this end, further approaches have been explored in creating functional nanodevices, such as molecular motors and nanoscale actuators [68, 69].

*3.2. Optical actuation*

Optical actuation for droplet motion relies on photosensitive/photoresponsive materials, such as azobenzene, spiropyran, cinnamate, *etc.*, which can be used to create surface gradients that lead to thermocapillary and optoelectrowetting effects, and Laplace pressure gradients [70–72], which can both be combined using various substrate patterns on a photopolymer surface during irradiation [71]. For example, asymmetric irradiation with UV-A light can lead to polar groups at exposed areas and, in turn, increased wettability. Then, the motion is expected to take place towards the more wettable regions. This concept can be nicely illustrated in the case of azobenzene, which is well-known for its trans– cis isomerisation and reaction upon UV light. Azobenzene can be used as a monolayer to alter surface properties and create a wettability (contact angle) imbalance that can drive the droplet motion [70], as illustrated in Figure 4. Asymmetrical irradiation leads to structural changes that create surface tension gradients, capable of driving the droplet towards one direction. Interestingly, by varying the direction of the photo-irradiation an analogous change in the direction of the movement is observed. In particular, the droplet moves in the direction of lower energy with a dynamic angle that is smaller than the angle $\theta_B$ and larger than $\theta_A$ (Figure 4). Such wettability gradients can be realised in both hydrophobic and hydrophilic substrates and can be exploited in lab-on-a-chip applications and optically reconfigurable microfluidics systems [72].

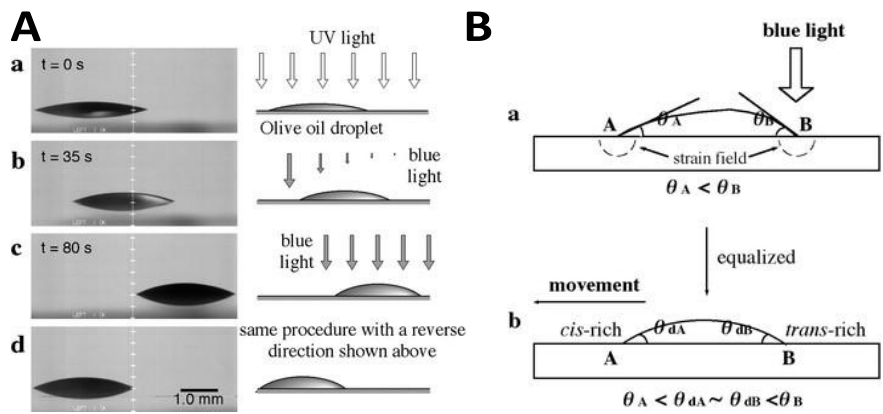

Figure 4: (A) Photographs at subsequent times light-driven motion of an olive oil droplet on a silica substrate, which has been modified with H-CRA-CM. As a result of asymmetrical irradiation with 436nm light perpendicular to the substrate, the olive oil droplet on a *cis-rich* surface moved in the direction of higher surface energy. The sessile contact angles changed from $18^0$ to $25^0$ (a, b, c). Changing the direction of the photoirradiation, the droplet was able to move in the other direction (d); (B) Cross-section of a liquid droplet on a surface with a surface energy gradient. In the absence of net spreading tension, the droplet may not move (a). In the presence of force imbalance, due to surface energy differences, the droplet moves with a dynamic contact angle $\theta_d$ being smaller than $\theta_B$ and larger than $\theta_A$ (b). Adapted from Ref. [70] with permission from the Royal Society of Chemistry.

*3.3. Mechanical actuation*

We now turn our attention to droplet motion induced by mechanical actuation, in other words, caused by mechanical vibration, acoustic waves, *etc*. Mechanical actuation often requires an asymmetric surface pattern or other gradients (*e.g.* a wettability gradient) to initiate and sustain the droplet motion. In this case, the most common patterns are ratchet-type surfaces [43, 44], where ratchets periodically repeat along the substrate. Such a saw-tooth-shaped topographyis combined with mechanical vibrations, which usually take place in the vertical direction. MD [74] and finite-element-method simulations [75] have explored how the vibrational energy can be transformed into direction motion in these systems.

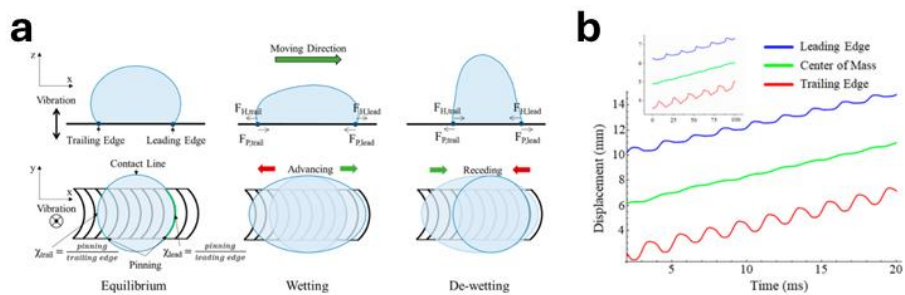

Figure 5: (a) The Anisotropic Ratchet Conveyor (ARC) uses vibrations to force the contact line of a droplet to oscillate through wetting and dewetting phases. By creating a difference in pinning forces between the leading and trailing edges, the surface's asymmetric pattern generates a net directional movement with every vibration; (b) Simulation of ARC transport can reproduce the experimental data. The lateral displacement of the simulated droplet at its centre of mass (green), as well as the leading (blue) and trailing (red) edges matches the behavior of the transports droplet in experiment shown as inset. Adapted with permission from Ref. [73]. Copyright 2017 American Chemical Society.

Figure 5 illustrates the mechanism of droplet motion on an anisotropic ratchet-type substrate, which undergoes sinusoidal, vertical vibrations in the experiment [73]. The asymmetry in the pattern, which repeats periodically along the substrate, leads to lateral motion of droplet of different volumes due to the resulting asymmetry in the forces as indicated in Figure 5a. Although leading and trailing edges follow an oscillatory pattern that creates an imbalance due to difference in the pinning forces between the leading and trailing edges as a result of the asymmetric surface pattern, the centre of mass of the droplet follows an almost linear displacement over time. This matches the behaviour of a transported droplet as shown in Figure 5b, where the vibrations have a smaller footprint on the centre-of-mass motion of the droplet.

Substrates can be submitted to both vertical and horizontal harmonic vibrations, which can be used to direct the fluid motion by adjusting the vibration frequency [46]. In this case, the motion resembles the ratchet-like mechanism, and the velocity of the droplet is tuneable and efficient enough to pull the droplet against an inclined substrate [43]. A numerical model demonstrating a ratchetlike motion due to vertical and horizontal vibrations and exploring the unbalanced capillary force, which arises at the contact line for hydrophobic substrates, has been built to study the underlying mechanisms and provides a range of capabilities of such systems achieving liquid droplet motion in preferable directions [76]. On the theoretical side, the combination of mechanical and electric actuation for a dielectric fluid by using the ratchet mechanism and switching periodically on and off the heterogeneous ratchet capacitor has been reported [45], where parallels between the self-ratcheting process and the class of thin film and Fokker–Planck equations have been drawn.

In the case of mechanical actuation combined with a wettability gradient, the interplay of mechanical and surface-tension forces can lead to enhanced or reduced droplet motion, depending on the direction of the mechanical force [77]. Moreover, ratchet-like setups can be combined with wettability gradients and mechanical or electric actuation achieving droplet transport [78]. Ratchet-like motion can also be induced by mechanical actuation that involves the stretching and relaxation of gradient substrates. This has been shown in the case of a droplet moving between two gradient surfaces compressed and extended periodically, where the droplet moves due to the wetting hysteresis [79]. This kind of mechanism might be relevant in digital microfluidics devices. Other ways of mechanically actuated droplet motion may refer to *torsion-triggered actuation* mechanisms [80]. This has been shown in the case of a graphene film. The formation of spiral wrinkles with shear deformations of smooth gradients in atomic contact and curvature and the combination of varying van der Waals and elastic energy landscapes are part of a mechanism that promotes the nanoscale droplet motion along the wrinkle troughs. As we will discuss later, however, wavy or wrinkled substrates are particularly relevant in the context of passive transport of droplets, a phenomenon known as *rugotaxis* [81], which has been investigated by MD simulation [11]. Further studies by MD simulations of graphene nanoflakes on graphene surfaces have highlighted a superlubricating directional motion that exploits the vibrational energy of graphene at various vibrational frequencies, where the latter play a key role in the directional motion [82].

*Slippery liquid-infused porous surfaces* (SLIPS), which will also be discussed in an upcoming section in the context of passive transport, have been used with high-frequency surface acoustic waves (SAW) as a mechanical way of causing droplet motion [83]. The combination of mechanical control and the lubricant layer leads to a more energy-efficient transfer from the waves to the droplet. Thus, significant droplet mobility can be achieved with less energy input, since the lubricant reduces pinning effects of the droplet at the contact line. Ultrasonic waves to control the dynamic transport of droplets on an aluminium plate have recently been reported [12]. However, in this case, the droplet undergoes deformations that in effect demand more energy than the directional motion [84]. Analysis of the results has determined the optimum conditions for minimizing droplet spreading and maximizing the droplet transport ability, which could be exploited in multifunctional droplet manipulation techniques. Furthermore, sonic excitations have been utilised to initiate droplet oscillations over a hydrophobic flat substrate and a mesh substrate (same contact angle of droplet) for various screen aperture ratios [85]. The droplet displacement was shown to positively correlate with an increasing sonic excitation frequency with vertical displacements being higher than the horizontal ones, while the maximum droplet displacement on flat substrates is higher.

A notable example based on acoustic actuation are *tactoids*, that is, droplets with a spindle-like shape, where the nematic and isotropic phases coexist as a dispersion of microscopic droplets. Dynamic behaviour of nematic tactoids has been observed as a result of the application of an acoustic field, with the tactoids exhibiting a spectrum of behaviour, such as stretching, bending, splitting, merging, spontaneous rotation, as well as locomotion [86]. In the case of droplet locomotion, tactoids offer various possibilities for active droplet control

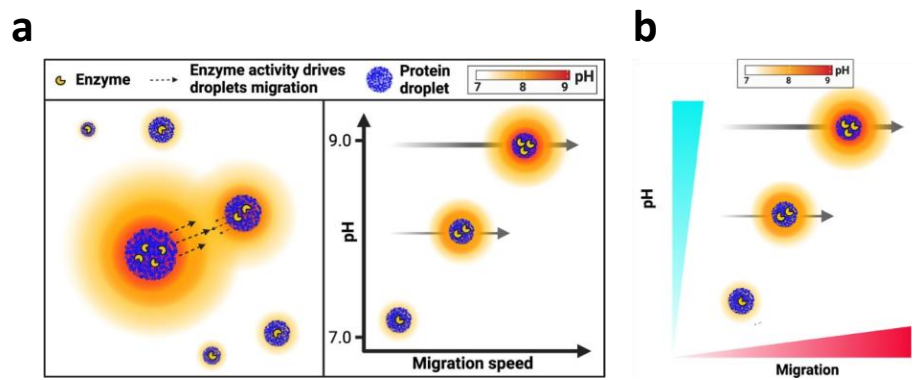

Figure 6: (a) Enzymatically active cell-sized droplets move freely by using internal enzymes to create a pH gradient outside their boundary. This external gradient drives and directs the droplets to migrate specifically toward their neighbours; (b) By adjusting the strength of the pH gradient, the modulation of droplet migration can be controlled. Adapted with permission from Ref. [38]. Copyright 2024 American Chemical Society.

and transport, due to both their response to external fields and changes in their structure.

*3.4. Active droplets*

Droplets can themselves induce changes to their immediate environment, which can lead to directional self-propelled transport, as, for example, in the case of droplets laden with surface-active agents [87]. The droplets move on a hydrophilic substrate with each droplet leaving behind a hydrophobic trace, which is attributed to the surfactant adsorption on the surface. Droplets are able to move uphill and exhibit a self-avoiding behaviour, since they will continue to move along hydrophilic paths, thus avoiding crossing the hydrophobic trails that have already been formed as a result of the droplet motion. Other artificial systems, such as chemotactic droplets, can respond to chemical gradients, providing mobility to otherwise immobile cells through bio-augmentation [29]. In particular, if cell lines produce biosurfactants only when inside hydrogel capsules, these surfactants can create a chemical interface to enable the droplet motion and the transport of cells.

A natural next step of this concept is the creation of enzymatically active, cell-sized droplets, which can move freely and interact with each other in a collective way [38].

An external *pH*-gradient created by the enzyme acts as a chemical signal and triggers droplet's migration towards neighbouring droplets. A strong correlation between enzyme's activity, which can be adjusted, and motion's efficiency in terms of droplet speed has been identified. Moreover, regulation of the *pH* offers further control on the effective migration of the droplet. These concepts have been explored recently [38] and are schematically demonstrated in Figure 6. Thus, the observed phenomena can in practice replicate processes, such as chemotaxis [88] and collective organisation, which can be proven useful to facilitate simple metabolic pathways, relevant in the area of synthetic biology.

Active fluids that can convert energy into mechanical work, causing them to flow as a whole or exhibit complex behaviour without external intervention, have been found in both biological and synthetic contexts. Notable examples here in the biological context are bacterial suspensions, where collective motion can create large-scale fluid flow and swirling patterns [89], sperm cells, where collective motion in a fluid creates self-organized flow leading to complex turbulent-like fluid behaviour, and the mixture of actin filaments and myosin motor proteins [90], which is essential for processes like cell division and motility. In the case of synthetic active fluids, notable examples refer to Janus particles [91], bimetallic nanorods, and surfactant-laden droplets [92]. More research in this direction is necessary to explore further possibilities of using active droplets in a controlled way along specific trajectories. This also entails modelling challenges on the theoretical side, due to the complexity of active systems leading to involved interactions between the components, as well as a large number of possibilities and environmental factors that may affect active systems. Following our discussion on droplet actuation and active droplets, we will now focus on cases of passive droplet transport.

*3.5. Wettability gradients*

Passive droplet transport relies on intrinsic surface properties, gradients, or spontaneous phenomena that cause the fluid to move without the requirement of a continuous external stimulus. The number of possibilities for passive transport is quite large and the distinction is made based on substrate properties that vary along a distance as expressed by wettability and topographic gradients (surface-tension gradients), Laplace pressure gradients, and others. As in the case of active transport, more than one system's property may contribute to the motion, and, moreover, some cases may combine active and passive processes to maintain the droplet motion.

Wettability gradients to trigger and sustain the droplet motion is a broad topic in the literature and encompasses the whole spectrum of research approaches, namely, experiment, theory, and simulation. For example, asymptotic theory has been employed to explain the droplet motion on a wettability gradient [93]. By matching properties at micro- and macro-scales, droplet's shape, velocity, and volume as a function of the substrate wettability can be determined [94]. From the perspective of MD simulations, the self-propelled motion of nanodroplets has been investigated for different types of wettability gradient, such as single gradient, continuous gradient, and nonlinear complex wetting gradient [95]. In the case of corrosive substrates of iron nanolayers, wettability gradients have been created by using different wettability pits under the droplet, which can lead to non-radial spreading and chemically active wetting gradients that can cause droplet motion [96]. Experimentally, highly adherent, gradient wettability patterns on superhydrohobic surfaces have been created using photolithography [97]. These have enabled the successful control of droplet motion on custom channels, offering possibilities for pump-free droplet transportation in microfluidics applications. For example, liquid metal nanodroplets confined within two-plate microchannels have shown unidirectional self-actuated transport through a continuous wetting gradient [98].

In the focus of various research efforts has also been the motion manipulation of *nano-objects* by means of gradient surface properties. In this case, the nano-objects are chosen to be flat, such as nano-flakes and nano-sheets, since they place the focus on the interfacial interactions. Two primary methods for causing the motion are through thermal gradient (thermotaxis) and via stiffness gradients (durotaxis). In particular, MD simulations have demonstrated that a thermal gradient on graphene surface can drive the motion of a smaller graphene nanoflake [99]. These findings have highlighted the role of the temperature gradient as the main driving force. MD simulations have also been carried out for droplets on self-assembled monolayers under different gradient temperature conditions [100]. In this case, it has been found that both the temperature field and the interfacial wettability determine the droplet motion, with droplets on hydrophilic surfaces migrating towards the cooler regions, while those on hydrophobic surfaces migrated towards the warmer regions for a given temperature gradient. Moreover, higher temperature gradients lead to higher droplet speeds. Following these studies, the durotaxis motion has been explored with the minimisation of the van der Waals potential energy being identified as the driving force for the directional motion of the nano-objects [37, 101, 102]. The same concept applies in the case of thermotaxis with droplets 'sitting' on cooler regions establishing a firmer contact with the substrate (lower interfacial energy). This type of passive transport onto elastic membranes has been explored for nanoparticles via MD simulations with motion up or down the rigidity gradient, depending on the specific mechanical properties of the membrane [103]. Such processes will be discussed in more detail in the following.

Specifically, in our MD studies, the mechanisms of the motion of a liquid droplet along stiffness gradients have been investigated [10]. These stiffness gradients also change the effective wettability of the substrate, since the droplet in the stiffer substrate regions has a smaller contact angle, as is clearly shown in the example of Figure 7a. In addition, the affinity of the droplet to the substrate can be modulated by the strength of the Lennard-Jones interactions between the substrate and the droplet with a range of possibilities, that is, varying from less to more wettable substrates. For this system it has been found that the mean velocity of the droplet during the durotaxis motion correlates positively with an increasing stiffness gradient and wettability. Moreover, smaller droplets were found to exhibit more efficient motion in terms of droplet speed and success rate of reaching the stiffest end of the substrate.

Since the motion is caused by differences in the interfacial energy between the droplet and the substrate, which is the driving force according to the mechanism of Figure 1a, and thermal fluctuations play a role, the imposed linear gradient neither leads to a linear change in the interfacial energy between the droplet and the substrate nor to an instantaneous velocity that depends linearly on the gradient during the motion [10]. Interestingly, by monitoring the velocity field of the droplets during the durotaxis motion, no indication of a rotating or carpet motion has been identified for these nanodroplets.

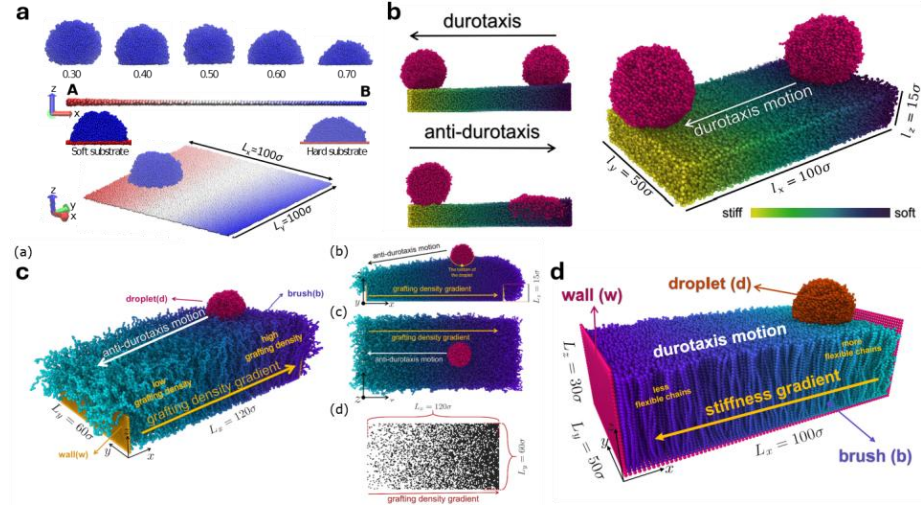

Figure 7: (a) Durotaxis motion towards stiffer areas of a surface of a droplet on a solid substrate with a stiffness gradient in the *x* direction. The contact angle of the droplet depends on the attraction force between the droplet and the substrates, as well as the substrate stiffness. Droplets on softer areas have a higher static contact angle, while when 'sitting' on the stiffer parts the static contact angle is smaller. Adapted from Ref. [10]; (b) Gel substrate with stiffness gradient that can cause durotaxis and antidurotaxis motion. The direction of motion depends on the composition of the liquid droplet. Droplets that highly wet the substrate will perform antidurotaxis motion. Droplets that wet less the substrate will move through the durotaxis mechanism [15]. Figure adapted from Ref. [15]. The antidurotaxis and durotaxis mechanisms can be demonstrated in the case of computer designed brush substrates where the droplet performs (c) antidurotaxis [14] (Figure adapted from Ref. [14]) and (d) durotaxis (Figure adapted from Ref. [13]) motion. The stiffness gradient in the case of brush in panel (c) relies on modifying the grafting density of the brush, while that in panel (d) for durotaxis motion relies on modifying the stiffness of the individual polymer chains along the gradient direction.

*3.6. Topographic gradients*

The concepts of wettability and topographic gradients are closely related, since topographic gradients tend to alter locally the wettability (equilibrium contact angle) of the substrates. In particular, topography can control the interfacial area between the liquid and the solid substrate, thus modulating the local strength of interactions, with various examples provided in a recent review [104]. The combination of wettability gradients and surface topography may offer more precise and efficient droplet motion in terms of speed, stability, and distance, which is particularly relevant for applications, such as self-cleaning and water harvesting.

Not only are ratchet-like surfaces relevant for mechanical actuation, but, moreover, combined with wetting surface gradients, can lead to directional liquid transport processes that can be observed in both natural phenomena and engineering applications. A characteristic showcase of nature is the capillary ratchet in shorebirds, which is a remarkable example of a feeding process, which is crucial for survival of life [105]. The capillary ratchet mechanism is used to transport the prey from the tip of the beak to their mouth. The characteristic geometry of the beak and the dynamics of its motion are optimized for efficient transport, with the wetting properties of the beak surface playing an important role in the transport. Unfortunately, changes in the wetting properties, *e.g.* due to the presence of pollutants, can impair the bird's feeding strategy.

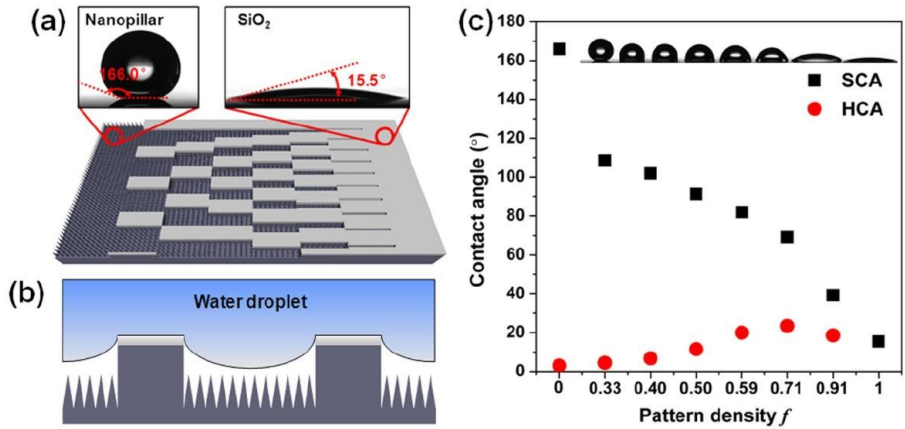

Figure 8: Surface properties and wettability characterisation. (a) A top-down schematic of the gradient surface, where the density of 1 mm × 90 $\mu$m × 2 $\mu$m SiO2 stripes increases from left to right. The pattern densities are 0, 0.33, 0.40, 0.50, 0.59, 0.71, 0.91, and 1, with inter-stripe spacing decreasing from 180 $\mu$m to 0 $\mu$m. (b) The undercut structure observed at the solid–liquid interface. (c) The static contact angle (SCA) and hysteresis contact angle (HCA) measured for a water droplet on each section of the gradient surface. Reproduced from Ref. [107].

In another example related to an engineering application that is based on biomimetic ratchet structures, a phase-field Lattice Boltzmann (LB) method has been used to simulate the motion of droplets on such surfaces [106]. Combining this method with analytical calculations, it has been demonstrated that special ratchet-type surfaces can be designed that can cause droplet motion with various factors affecting the liquid transportation, such as droplet deformation and pinning/depinning of the contact line. A force balance analysis can provide further details on the underlying mechanisms of such droplet motion on a ratchet structure, as illustrated in Figure 5, thus providing further guidance for the design of microchannels for intelligent fluid transport.

The synergistic combination of wedge-shaped tracks (*topography*) and wettability gradients has attracted attention because it offers high peak mean velocities, despite a slower initial acceleration compared to an individual gradient of one or the other type in this case [108]. Such substrates have also been studied by MD simulations [109]. In addition, wedge-shaped surfaces with alternating hydrophilic and hydrophobic regions have shown that the movement of water droplet towards hydrophilic regions is more efficient than in the case of non-alternating surfaces [110]. Moreover, the use of simultaneously decreasing pillar width and spacing can reduce advancing and receding free-energy barriers, resulting in lower friction properties, particularly relevant for applications in microfluidics [111]. The droplet motion here is driven by the roughness gradient, while the underlying mechanisms have been discussed in terms of thermodynamic arguments. Although such a motion has been reported in the case of hydrophobic substrates, designs of wettability gradient surfaces for both superhydrophobic and hydrophilic cases have been proposed to transport droplets at longer distances by structural topography [107].

This possibility, which provides a more general design of similar setups to induce wettability gradients, is presented in Figure 8 [107]. Typically, the pattern densities gradually increase in one direction, which results in changes in the static contact angle, that is, a wettability gradient is created as a result of changes in the topography. As shown in Figure 8, the hysteresis contact angle, which is responsible for the force imbalance driving the movement of the droplets, increases with the density of the pattern, $f$, reaching a maximum value at about $f = 0.71$, while the static contact angle decreases monotonically. In addition, theoretical calculations based on force analysis can be employed to predict the maximum displacement with possibilities considering the motion of droplets of different volume along arbitrary paths.

Gradient nanostructures can be created experimentally through various techniques, such as non-uniform interference lithography [112] and scalable bladecut masking techniques [113]. In the former case, the use of Gaussian-shaped intensity distribution of two coherent laser beams eliminates the need for expensive etching and e-beam evaporation during the mould fabrication. Then, the mould is used to reproduce nanostructures with gradient wettability, which can also be studied theoretically. In this case, topography-patterned wettability contrast on aluminum surfaces was created, which is suitable for thermal applications, achieving significant water contact-angle contrast for enhanced droplet shedding on superhydrophobic regions during condensation. Thus, dual-length scale microstructures can be created to reduce resistance to droplet motion by preventing direct contact between the droplet and the solid [114]. Previous methods based on wettability contrast are generally limited by the resistance to motion, which is avoided by creating these dual-length microstructures. Such surfaces also exhibit high adhesion, which makes them suitable for applications such as fog harvesting and heat transfer. Recently, a biomimetic system that exploits hybrid and wettability gradients has been manufactured by means of laser etching and gradient anodising technologies [115]. The combined effect of the wetting driving force and gravity has been shown to accelerate the discharge of the droplets for fog collection, exhibiting an efficiency that is 80% higher compared to a reference system without the gradient.

Controlling oil droplet motion on oleophobic surfaces has been performed on a number of occasions, despite challenges due to their low surface tension. In particular, oil droplet self-propulsion has been reported in hydrophilic surface microtextures with radial arrays of undercut stripes [116]. Three modes of oil motion (inward transport, pinned, and outward spreading) have been reported, which can be controlled by structural parameters, such as the intersection angle and width of the stripe. Moreover, a theoretical model can provide insight into the structure–droplet-motion relationship. The issue of oil droplet manipulations has also been discussed in the case of fabricated pillar arrays with mushroom-shaped profiles (increased *roughness*) to create oleophobic surfaces with angles higher than 170° for hexadecane droplets [117]. Local energy gradients are created through textured surface ratchet tracks with arc-shaped pillars that achieve speeds of up to 7 mm/s.

Topography changes relying on *curvature gradients*, which may refer to both patterned and curved substrates, such as wrinkled substrates with gradient in wavelength characterising their wavy structure, can lead to a self-sustained droplet motion. In the case of curvature-gradient substrates, it has been shown that they can spontaneously propel micro-droplets at relatively high speeds [118]. A similar principle has been explored in the biological realm to demonstrate in a minimal cell model how cells can migrate on curved surfaces [119]. The specific term used in this case is *curvotaxis*, where the motion is presumably driven by the minimization of adhesion and binding energies between the cell's membrane and the substrate's curved surface. A common theme is based on the fact that the shape and curvature of the underlying surface are the primary drivers of motion, which can be tuned by varying surface properties. In this case, it would be interesting to explore the role of the elastic energy of a droplet during transport.

Combining different surface features, such as Janus pillars, namely half-flat and half-curved surfaces, has been used to cause droplets to move along the curved side of substrates by a geometry-driven effect [91]. This effect can be scaled up to achieve continuous long-range transport of water collected from fog. Similarly, MD simulations have shown how nanopillared surfaces affect droplet motion based on a rolling mechanism of droplets in the Cassie–Baxter state [120]. These studies demonstrate how different topographic and chemical properties of the substrate can be used to control droplet motion in technologies that are relevant for microfluidics and anti-icing coatings.

Patterned substrates with surface topography such as wrinkles and grooves with gradient properties have been explored in various studies. In the case of wrinkled substrates [11, 81, 122], the motion is known as *rugotaxis* and the droplet moves typically due to a gradient in the wrinkle wavelength towards regions with a higher density of wrinkles, thus minimising (more negative) the interfacial energy between the droplet and the substrate. A typical system as studied by means of MD simulation for nanodroplets on wrinkled substrates with a wavelength gradient of the wrinkles is presented in Figure 9a [11]. Understanding rugotaxis motion at nanoscales requires a detailed analysis of droplet's state, that is, whether the droplet is in a Cassie-Baxter or a Wenzel state. In this case, theory can accurately predict the state of the droplet on wavy surfaces without gradient [121], and these predictions for substrates with different wrinkle's wavelength and amplitude have been corroborated by the MD simulations (Figure 9b,c) [11]. This is relevant as pinning effects, which may hinder the droplet motion, are more common when the droplets are in the Wenzel state, that is, the droplet fully wets the wrinkles. Further theoretical analysis of rugotaxis can be realized by means of classical density functional theory (DFT), where one can obtain the local density of the system at each point in space directly from MD simulations, and the various free energy components can be analyzed individually. In this way, the dominant role of the liquid–solid interfacial energy in driving the spontaneous droplet motion has been confirmed [12]. During the minimisation of the energy the mechanisms described in Figure 1 continue to hold and the imbalance in the surface-tension force (integrated over a distance [6]) is responsible for the droplet motion.

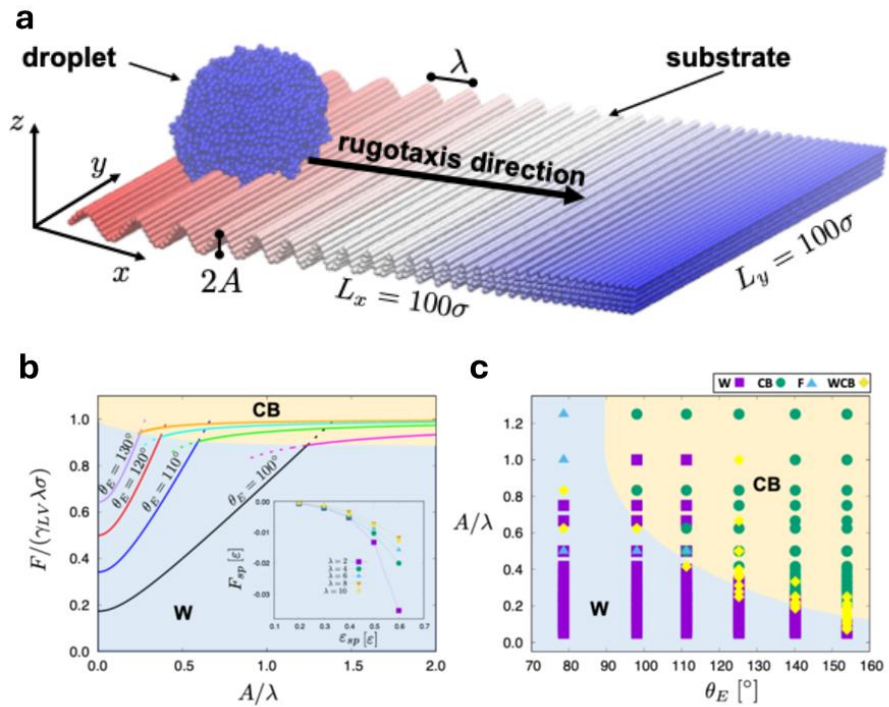

Figure 9: (a) Typical system of a droplet that can perform rugotaxis motion towards wrinkles with smaller wavelength simulated by MD simulation. (b) Theoretical prediction of the free energy as a function of the surface roughness (amplitude, $A$, over wavelength, $\lambda$ [121]. Inset shows the interfacial free energy as a function of the droplet–substrate attraction parameter for different wrinkles wavelength. (c) Simulation predictions for Cassie–Baxter (CB) and Wenzel states (W) *versus* theoretical predictions. Vertical axis for surface roughness, while $\theta_E$ is the equilibrium contact angle of the droplet on a flat surface. Figure adapted from Ref. [11].

In the case of circular groove arrays, impacting droplets have been shown to bounce and be transported in a specific direction [123, 124]. The size and direction of the wrinkles, as well as the contact angle can affect the rugotaxis motion as has been shown experimentally [81, 125]. Further experimental work on soft pre-strained polydimethylsiloxane (PDMS) substrates has shown droplet motion due to the gradient of the wrinkle wavelength, which helps understand the impingement of the droplet on the gradient substrate towards applications of droplet transport along a certain direction [126]. While pinning effects might hinder rugotaxis droplet movement[11], in other cases, pinning and depinning effects combined with a wettability difference between two surfaces can be exploited to control droplet movement. Here, the physical pinning barrier can be overcome and the time to cross the barrier by the droplet can be tuned, as has been shown by MD simulations [12]. Moreover, different pinning barriers can allow droplets of a certain size to overcome the barrier, thus allowing a natural selection of droplets to cross the physical barrier depending on their size. Such system designs might be of interest in applications such as digital microfluidics.

Gradients in surface chemistry or topography lead to asymmetric forces at the contact line of the droplet, and in turn to self-propelled motion, as has been described (Figure 1) [6]. On soft substrates, this driving force is particularly relevant as the material deforms at the contact line, where the force imbalance can be described by the Neumann triangle model [127]. Although this imbalance can be viewed as the difference between the droplet's advancing and receding contact angle during motion, this distinction is challenging to observe in nanoscale simulations, for example, by using MD simulations. Hence, from a molecular point of view, the minimization of the solid–liquid interfacial energy, which incorporates the imbalance of the uncompensated Young forces integrated over a distance along the motion, appears to be a suitable criterion for exploring the directional self-propelled motion [10, 11]. The way that the system minimizes its interfacial energy in various brush and gel substrates will be the focus of the following discussion [13–15].

Experiments of water droplets on a silicon gel substrate with stiffness gradient have demonstrated the self-propelled directional droplet motion at micron scales towards softer regions of the substrate [33]. A model of a droplet on a viscoelastic, soft substrate of varying thickness moving towards thicker, stiffer regions has investigated the asymmetric deformation at the contact line, which drives the motion [128]. Further exploration of this concept has indicated that a stiffness gradient alone may not provide the most efficient way of transport [129]. More recent experimental studies have shown that millimetre-sized droplets move along a cross-linking density gradient on soft gels, with the driving force being able to drive droplets uphill against gravity [130]. In this case, surface energy difference due to variations in the cross-linking density provides a new strategy for controlling droplet dynamics on soft, dissipative substrates. In practice, this difference in cross-linking density results in a variation of stiffness, as has been recently shown by means of MD simulations for a gel system that is based on a similar concept for the gradient (Figure 7b) [15]. A hydrophilic capillary bridge on a soft surface would induce motion in the opposite direction of a sessile droplet, due to the interplay between soft layer deformation and contact-angle modulation[131]. Specifically, the wetting ridge rotation aligns with Laplace pressure, and a hydrophilic bridge on a softer layer can lead to an increased contact angle and movement from thicker to thinner areas, opposite to sessile droplets.

Gel substrates with a stiffness gradient have been explored using MD simulations for the motion of a liquid droplet towards the softer or the stiffer regions of the substrate, depending on the degree that the liquid can wet the substrate (Figure 7) [15]. Hence, this is one of very few systems in the literature that have demonstrated bidirectional droplet motion on a substrate with stiffness gradient. In particular, the stiffness gradient was achieved by varying the degree of cross-linking density between the polymer chains, similar to recent experiments discussed above [130]. Droplets with a small equilibrium contact angle move to the softer parts, that is, regions with lower cross-linking density, while droplets with a larger contact angle move towards the stiffer areas [15]. The directional motion in these two scenarios is characterized by different mechanisms that minimise the interfacial energy, which have separately been identified and investigated in detail in two different designs of polymer brush substrates (Figure 7) [13, 14].

In the first brush design (Figure 7d) [13], the stiffness gradient is achieved by varying the individual stiffness of the grafted polymer chains along a specific direction, which in turn results in a stiffness gradient along the brush substrate. The durotaxial motion takes place in the direction of the stiffness gradient, namely from the softer to the stiffer regions. The latter is characterized by a lower roughness, which results in a higher number of molecular contacts between the brush chains and the droplet. Hence, this state is characterized by a lower interfacial energy and drives the durotaxis motion.

In contrast, in the second brush substrate design (Figure 7c) [14], the stiffness gradient was implemented by varying the grafting density of the brush substrate, which consisted only of fully flexible polymer chains. This resulted in a stiffness gradient in which regions with a higher grafting density had a higher stiffness, while regions with a lower grafting density of the brush had a lower stiffness. For this substrate, the motion takes place from the stiffer towards the softer parts of the substrate (antidurotaxis). Again, as the droplet moves along the substrate, the interfacial energy between the droplet and the brush substrate becomes lower. However, in contrast to the first brush design [13], the energy minimisation here is due to the penetration of the droplet into the brush substrate as it moves towards the softer parts (Figure 7c). These two distinct mechanisms of durotaxis and antidurotaxis motions are observed in the case of the gel substrate with stiffness gradient [15]. In a durotaxis motion, the droplet also performs a diffusive motion in the $y$-direction perpendicular to the *x*-axis (direction of the gradient), whereas in the case of antidurotaxis the motion mostly takes place along a certain direction, opposite to the gradient, with minimal deviations, indicating the greater role of interfacial interactions in determining the droplet motion.

It remains to be seen whether these substrates designed by simulations [13– 15], will be experimentally realized in the future. Moreover, it will be interesting to see how topological changes in droplets (*e.g., droplet coalescence*) are affected by the presence of the gradient substrate, especially in the case of soft substrates, where a longer range of domain influence is expected than the length of the intermolecular interactions [14].

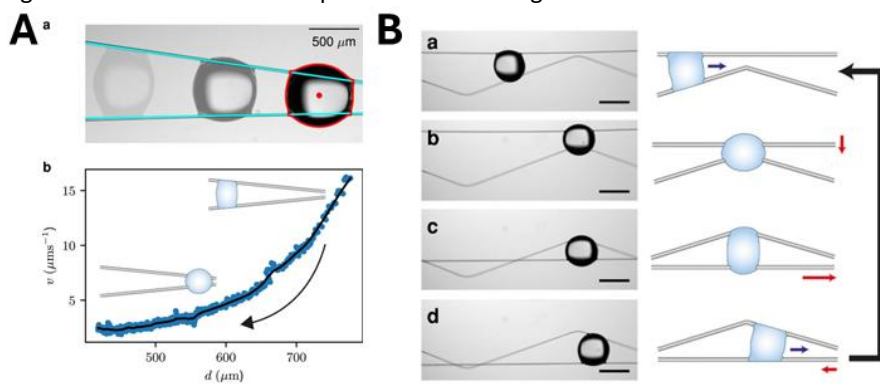

Figure 10: (A) Plot of droplet speed as a function of the distance between the fibres at the position of the droplet centre. We note that the droplet moves towards the apex, thus from the left (large *d*) to right (small *d*) in this plot (sketches of typical configurations are inset). (B) Images and corresponding sketches of droplet migration driven by the ratchet mechanism (scale-bars are $500\mu m$. a) Droplet moving toward the apex. b) Droplet at the corner of a sawtooth, once at this point the straight fibre can be translated downwards to the other size of the sawtooth fibre. c) Droplet at point of unstable equilibrium, the straight fibre can be translated horizontally to initiate droplet motion in the desired direction. d) Droplet once again moving spontaneously toward the apex, at this step the straight fibre can be retracted horizontally to its original position. Adapted from Ref. [132], licensed under a Creative Commons Attribution-NonCommercial (CC BY-NC-ND) 4.0 license.

Microscopic droplets placed between fibres held at a fixed angle can perform directed fluid motion [132]. This mechanism, which resembles the way shorebirds are fed through their beak [105], is illustrated in Figure 10. In contrast to the beak, system's design here is based on fibres with a varying distance between each other. Considering the distance between the fibres at the position of the droplet centre, this has shown a positive correlation with the transport speed of the droplet. Moreover, the droplet motion can induce changes in the structure of the fibres, in repeated cycles of a ratchet-like mechanism, which involves specific steps during a repeatable cycle, that is, droplet motion towards the apex, translation of the straight fibre downwards to the other side of the sawtooth fibre, once again spontaneous motion towards the apex, and horizontal retraction of the fibre to its original position, as shown in Figure 10.

Based on a similar mechanism, non-parallel deformable channels can be used to precisely control the direction of droplet motion [133]. Parameters such as the initial channel configuration and its flexibility, can be used to influence the motion. Related examples from nature here refer to wings of a specific butterfly, where a ratchet-like micro/nano structure of the wings leads to an asymmetric growth and de-wetting of fog droplets (Figure 11) [3]. In turn, this imbalance leads to directional transport in both static and vibrating states of the wings. A similar case of artificial super-hydrophobic micro-tracks with converging micro-ridge topography is a design that leads to an asymmetric meniscus and a resulting Laplace pressure imbalance [134]. This has shown to be an effective way for long-distance droplet transport that can even lead to an uphill motion against gravity. Here, the mechanisms are related to the influence of geometry on pressure and surface energy. A model has been put forward to explain how a V-shape groove directs droplet motion, highlighting the critical role of the groove's geometry and wettability of its walls in the efficiency of the motion [135].

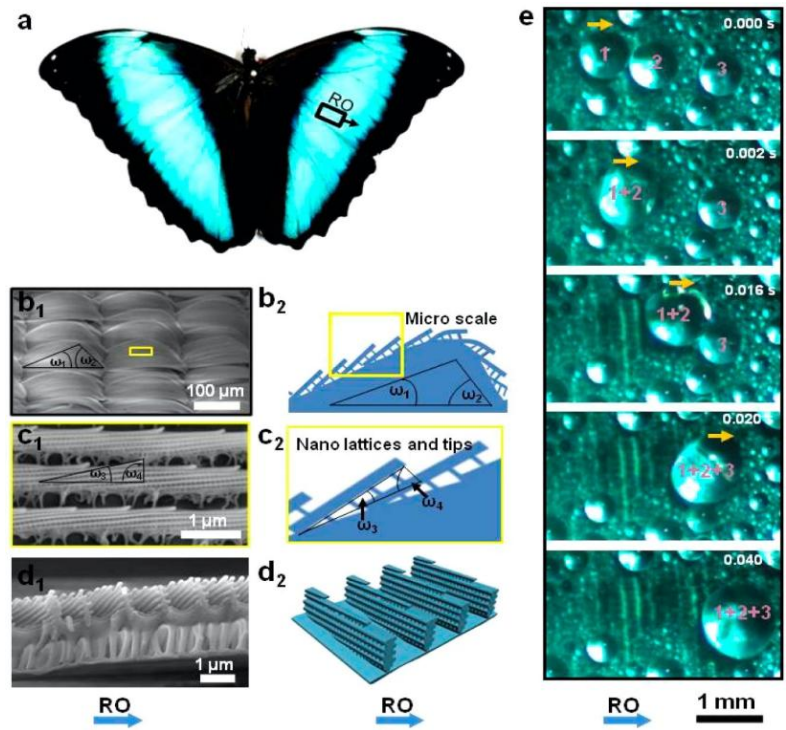

Figure 11: (a) An *Amorphodeidamia* butterfly with a black arrow indicating the radial outside (RO) direction of the wing. (b1-d1) Environmental scanning electron microscope (ESEM) images show the hierarchical structure of the wing, from overlapping scales (b1) to the ridges on a single scale (c1), and finally, a cross-section of the ridges (d1). (b2-d2) Corresponding illustrations detail the micro-/nano-ratchet-like structure. (e) A top-down view shows the directional movement of fog drops (yellow arrows) on a stationary butterfly wing at a video rate of 500 f/s. Reprinted with permission from Ref. [3]. Copyright 2014 American Chemical Society.

In addition to the mechanical and topographic gradients, variations in the topology of the surface chemistry and charge can be used to self-propel droplets. Here, an example is mechano-tuneable, micro-textured chemical gradients in elastomer films [137]. The surface texture can be adjusted to precisely control the droplet transport by mechanically deforming the film substrate, which highlights the possibility of combining a chemical gradient with a tuneable physical one. In another case, a rewritable surface charge density gradient can be printed on various substrates [136]. This provides a powerful platform for programming droplet motion characterized by high-velocity and long-range transport without external energy supply. Moreover, the internal crystalline structure of a material can create an effective surface charge-density gradient. A schematic illustration and the relevant experiments are presented in Figure 12. The efficiency of this process is further highlighted by the vertical climb of the droplet along a superamphiphobic surface decorated with a static charge distribution
[136].

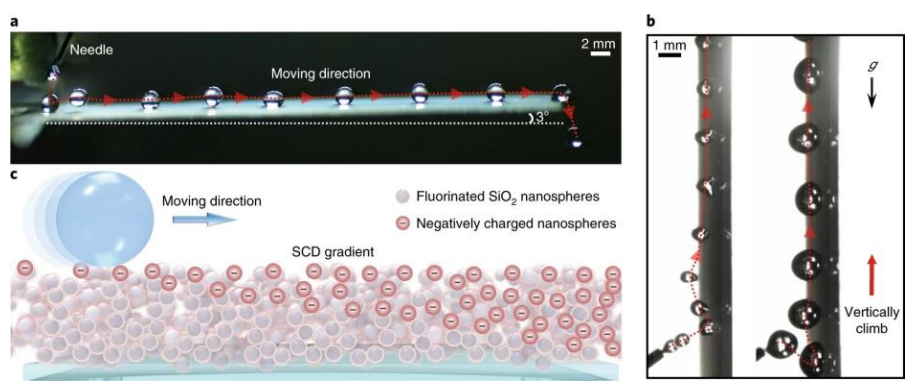

Figure 12: A printable Surface Charge Density (SCD) gradient that can mediate droplet transport. (a) Droplet position at different times on a superamphiphobic surface with an SCD gradient at a climbing angle of 3°. (b) Trajectory of droplets with different radii moving upward on a vertical superamphiphobic surface with an SCD gradient. (c) A schematic representation of droplet self-propulsion on a superamphiphobic surface with SCD gradient. Reprinted with permission from Ref. [136]. Copyright 2019 Springer Nature.

Piezoelectric single crystals have also demonstrated self-propelled droplet motion in specific pre-determined directions, as a result of anisotropic thermoelastic and piezoelectric interplay [138]. This is an example that shows that an intrinsic material property can create the necessary asymmetry for controlled droplet motion onto the surface. At nanoscales, torsion-triggered actuation on graphene has been shown to cause nanoscale transport [80]. Thus, even tiny forces at the atomic level can cause directional droplet transport. Flat, stretched, uniform soft substrates exhibit asymmetric wetting, which can be used to cause droplet motion [139]. The droplets are able to slide faster parallel to the applied stretch and appear elongated. This occurs due to droplet-induced deformations of the substrate near the contact line, similar to what has been observed for droplet oscillations on solid substrates [140]. In the case of a strain gradient, molecular transport on graphene has been achieved, which is induced by varying surface tension through stretching, thus demonstrating another possibility for directional transport [141].

*3.7. Laplace pressure gradients*

The pressure difference across a curved interface between two fluids, for example, a liquid and a gas, is caused by surface tension and is related to the curvature of the interface, a definition that is usually employed in a continuum picture. In a rather simple form, this can be expressed by the relation $\Delta P = \gamma_{lv}(1/R_1 + 1/R_2)$, where $\Delta P$ is the Laplace pressure difference, $\gamma_{lv}$ the surface tension, and $R_1$ and $R_2$ the principal radii of curvature of the interface. A pressure gradient can direct liquid from a high-pressure (high curvature) region to a low pressure (low curvature) region. This phenomenon has been observed in both natural and engineered systems related to passive liquid transport [142].

Laplace pressure gradients, which can act synergistically with other driving forces, have been used in the case of engineered systems for droplet self-propulsion on superhydrophobic surfaces [134]. A textured surface with converging micro-ridges causes an asymmetric deformation of a droplet's meniscus that leads to a Laplace pressure imbalance, which, in turn, causes the motion of the droplet. Similarly, a multi-gradient surface on a titanium alloy can be used for spontaneously moving a droplet uphill [143]. In this case, the geometry and wettability gradients have been combined with the synergistic effect producing a Laplace pressure force that causes the droplet motion. Taking cactus *Opuntia microdasys* as an example of a natural system for fog collection, which consists of conical spines and trichomes, a Laplace pressure gradient along with a surface-free energy gradient has been shown to lead to droplet motion that helps collect water droplets from fog [4]. Such natural structures and others such as those encountered in beetle backs and spider silks [142] can be a source of inspiration for the design of biomimetic materials, which can cause droplet motion and can be used for the collection of water in arid environments, as well as in other applications.

*3.8. Marangoni effects*

The Marangoni effect, or Marangoni flow, relates to the transfer of mass along an interface between two fluids due to a surface tension gradient. The surface tension gradient can be created by various processes, such as concentration differences of the surfactant along an interface (*chemocapillary* effect) and temperature variations (*thermocapillary* effect). The Marangoni effect manifests itself through fluid motion from the lower to higher surface tension regions, dragging the bulk fluid along with it. Various aspects of this effect can be exploited in self-propelling droplet motion and their manipulation on surfaces, which is important for a range of areas from microfluidics to biology.

Several studies have focused on temperature gradients that can lead to droplet motion. For example, the thermotaxis motion of a deformable droplet in a confined flow has been investigated, highlighting the connection between the temperature gradient and the velocity of the droplet [145]. In particular, a linearly decreasing temperature profile will enhance the droplet migration, which highlights the interplay between thermal and flow fields and their resulting impact on interfacial stress and viscous drag. Thermocapillary motion of small droplets along thin fibres has also been observed, where droplet motion takes place from warmer to cooler regions, which is theoretically modelled [144]. The mechanism of this motion and the flow field generated is illustrated in Figure 13 by repeated snapshots of the droplet as it moves along the fibre over time towards the cooler part. In particular, the fluid flow has a different profile depending on whether the observer is attached to the fibre or is moving with the centre of the droplet as shown in Figure 13. Similarly, an energy-conservative many-body dissipative particle dynamics model has been used to simulate the thermocapillary motion of a water droplet onto a hydrophobic surface [146], while another study has shown that droplets can move towards regions of lower energy due to a temperature gradient, given that surface tension is temperature-dependent [6]. In all of these cases, imbalances of surface tension manifest themselves through the Marangoni effect.

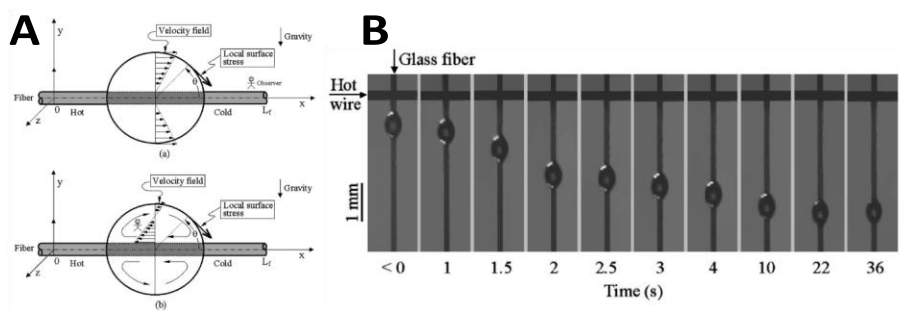

Figure 13: (A) A sketch of a spherical droplet on a fibre and experimental images of an n-decane droplet. The flow field from two perspectives: (a) an observer on the fibre and (b) an observer moving with the droplet's centre; (B) Images of a 50.22 mm n-decane droplet ($a_m = 50.22mm$, $l = 50.72mm$) at various times. The experiment was conducted with both the hot wire and glass fibre in a horizontal plane, as seen from a top view. Reproduced from Ref. [144], with the permission of AIP Publishing.

Chemical concentration and humidity gradients can be used to cause and manipulate droplet motion leading to Marangoni flow. For example, chemical gradients can be created on graphene surfaces, which can pull or push droplets depending on the composition of the gradient (*e.g.*, oxygen or fluorine groups) [148]. Also, subnanometer motion of cargoes driven by thermal gradients along carbon nanotubes has been reported [149]. In another case, stimulant gas can cause the motion of oil droplets in an aqueous phase by creating a spatial gradient in surface tension [150]. Droplets with chemicals can also react with a patterned surface, which in turn generates a sufficient surface-tension gradient that can propel the droplet [151]. LB simulations have explored a similar mechanism driven by humidity gradients, where droplets move towards regions of higher vapour concentration due to wettability differences and Marangoni flow resulting from surface tension variation [152]. Here, a representative example is the droplet motion of evaporating surfactant-laden droplets, which is caused by Marangoni vortices in the vicinity of the contact line (Figure 14) [147]. The interfacial flow takes place towards the back side of the droplet, where the surface tension is higher, with the generated Marangoni flow resulting in the droplet motion. The displacement of single tracked particles in the left, right, and middle parts of the droplets is shown in Figure 14, as well as schematics of the underlying mechanisms and Marangoni vortices. In a broader context, various studies have discussed methods for controlling droplet motion. This includes previous reviews on microfluidic actuation methods, such as those related to Marangoni stresses and thermocapillary flow on chemically patterned and textured surfaces [153].

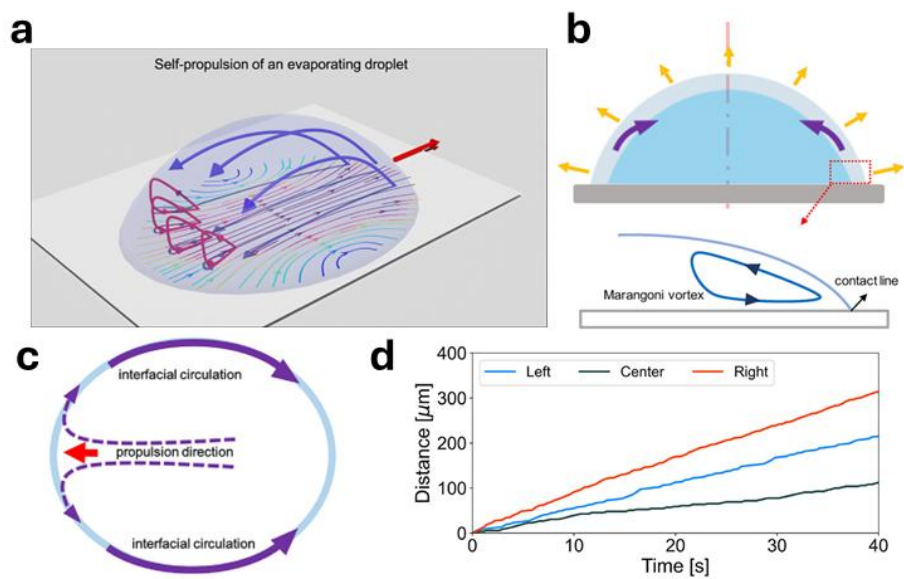

Figure 14: (a) Self-propulsion of an evepourating droplet on a polymer-coated substrate, which is driven by a surface tension gradient. (b) The mechanism behind droplet contraction in a surfactant-laden droplet is also shown, detailing how Marangoni vortices near the contact line cause this effect. (c) A schematic picture demonstrates the interfacial flow to the back of the droplet, where the surface tension is higher. (d) the plot compares the displacement of particles tracked at the centre and on the sides of the droplet. Adapted from Ref. [147].

*3.9. Slippery liquid-infused porous surfaces*

Slippery liquid-infused surfaces (SLIPS) have been inspired by the surface of *Nepenthes* pitcher plants, which offer self-healing liquid and ice repellence, pressure stability, and low contact angle hysteresis, which reduces the friction of the droplet motion [154]. Unlike superhydrophobic surfaces that depend on a stable air–liquid interface as a result of surface roughness to maintain a liquid phase in a Cassie–Baxter state, SLIPS are created by infusing a porous or textured solid substrate with a lubricating liquid. The formation of a lubricant film creates a smooth, defect-free surface onto which droplets of an immiscible liquid can move with low friction. This occurs because the film is highly repellent to various substances, such as water, oils, or even blood. Thus, various limitations of superhydrophobic substrates can be addressed, including failure under pressure or damage, as well as self-healing capabilities [154].

The droplet motion on SLIPS can be achieved by creating a surface gradient property. Several mechanisms have been explored, such as those that rely on topographic gradients, showing bidirectional self-propelled droplet motion, which was investigated by both LB simulations and experiments [155]. The direction of motion, whether it is towards the denser or sparser solid fraction, depends on the wettability difference between the droplet on the solid surface and the lubricant. In this case, the interplay between surface topography, lubricant, and droplet properties will eventually determine the direction of motion.

Lubricant-impregnated surfaces have demonstrated the ability to control and position droplets [156]. The patterning created menisci features in the lubricant layer, which interact with the droplets via a capillary mechanism that enabled controlled localization. An advanced approach is based on the creation of biphilic SLIPS. In this case, a substrate is patterned with two distinct lubricant-infused domains of a wedge-shaped pattern [157]. This creates a gradient that can propel the droplets at a long distance. The motion relies on the wettability difference between the lubricant domains, for which an analytical model has been developed to predict the motion based on both the wettability and the angle of the wedge's opening. This system is illustrated in Figure 15A, where a droplet is self-propelled on a composite two-lubricant oil SLIPS, with the transport taking place towards the spatial region dominated by an oil of higher wettability geometrically designed in such a way to create a wettability gradient. In particular, a droplet would self-propel from the apex towards the broader end of the wedge with a centre-of-mass velocity that quickly decreases. For this system setup, Young's equation still holds accounting for the additional interfacial forces that emerge when the lubricant cloaks the droplet as shown in Figure 15B.

Using SLIPS in practical applications requires fundamentally understanding their hydrodynamics and stability. A key challenge is the depletion of the lubricant film, which can compromise the surface's repellent properties, thus increasing the friction. Research has indicated that the droplet can entrain liquid with it as it moves, similarly to the Landau–Levich–Derjaguin problem, which leads to the formation of a wetting ridge around the droplet and depletes the lubricant [158]. In contrast to what might be expected, faster-moving droplets were found to deplete less lubricant than slower ones, an observation that holds important implications for durable and long-lasting SLIPS.

In a more complex scenario, four different phases are present when a droplet is placed on a SLIPS, that is, the droplet, the lubricant, the solid, and the surrounding air. Such a system can create multiple three-phase contact lines and a finite annular ridge of lubricant around the droplet [159]. These formations play a significant role in governing both the initial resistance to droplet motion (pinning) and the subsequent sliding velocity during the motion. Understanding the hydrodynamics of the associated processes help better control and predict the droplet mobility on SLIPS.

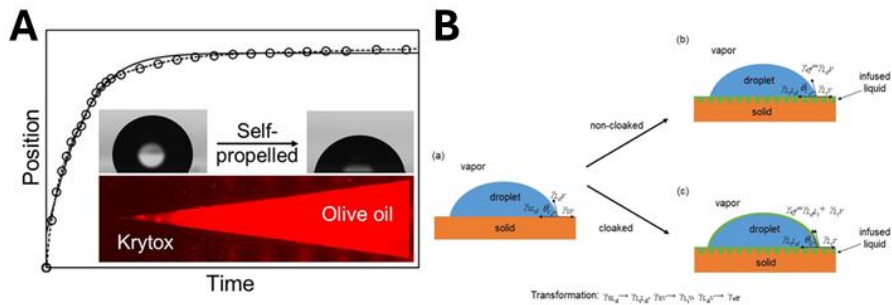

Figure 15: Droplet self-propulsion on a composite SLIPS wedge and relevant wetting laws. (A) Side views of a droplet at two points along its path as it moves from the apex toward its equilibrium position. In the fluorescence image, red indicates the more wettable olive oil regions, while black corresponds to the Krytox-coated regions. (B) Young's law and the liquid Young's law, with contact angles shown for a droplet resting on (a) a solid surface, (b) an infused-liquid film that does not cloak the droplet, and (c) an infused-liquid film that does cloak it. Adapted from Ref. [157]

Droplet motion that shares characteristics with SLIPS is the one based on the Leidenfrost effect, where a liquid droplet levitates on a cushion of its own vapour on a very hot surface (Leidenfrost point on a pool boiling curve [160]). This effect can be combined with ratchet-like textured surfaces to cause the droplet movement [161, 162]. More complex systems combining the Leidenfrost effect with a herringbone-ratchet surface to obtain low-friction leading to selfcentre ring droplet propulsion have also been investigated [163]. A study has further explored the possibilities of self-propelled motion on a heated surface of a ratchet-valley array (RVA), fabricated via laser micro/nano processing technology, in different boiling states, that is, nucleate, transitional, and film boiling [164]. While heat transfer processes are expected to significantly be influenced by the cavity geometry between the liquid and the solid [165, 166], the direction of motion was different for the various boiling regimes. Thus, this specific surface design has demonstrated bidirectional motion depending on the temperature of the substrate with the vapour bubbles developing underneath the droplet playing an important role in the force balance. In particular, in nucleate and transitional boiling regimes, the droplet moves along the RVA through two distinct mechanisms for each case, namely the adhesive and rebound modes, respectively. In contrast, in the film boiling regime the droplet was observed to move against the direction of the RVA. Similarly, a gradient boron nitride nano-sheet grid surface was shown to enable directional self-propelled motion below the Leidenfrost temperature by trapping air, which acted as a lubricating layer [167]. These examples demonstrate the possibility of vapour films acting as lubricating layers, which can be combined with a gradient property or an asymmetric pattern to lead to directional self-propelled motion. However, further discussion about such phenomena is beyond the scope of this review.

4. Anticipated applications and perspectives

Controlling droplet and cell motion precisely with the range of possibilities discussed above for gradient substrates naturally leads to an array of anticipated applications. In fact, bio-inspired surfaces have been implemented in biomimetic systems relying on hybrid and gradient wetting surfaces, which have been manufactured using laser etching and gradient anodising technologies [115]. Taking advantage of the synergistic effect of a wetting driving force and gravity, these systems have shown an 80% increase in efficiency in fog water collection compared to the control sample. Moreover, the device integrates droplet capture, coalescence, and transportation functions, thus providing a new approach for the design of efficient mist collectors, while having direct implications for microfluidics systems as well. In general, water harvesting relies on the development of efficient systems to collect water from fog or condensation. Many of the relevant processes have been inspired by natural structures such as cacti and butterfly wings (*e.g.* see Figure 11) [3, 4, 110].

Programmable droplet motion based on gradient substrates can be used in digital microfluidics, specifically platforms for chemical synthesis, theranostics, and lab-on-a-chip technologies. These technologies require the precise handling of liquids in processes such as pumping, mixing, micro-reactions, sorting, and programmed liquid delivery [5, 12, 53, 57, 62, 72, 106, 108]. In this regard, processes, such as separation and sorting of deformable entities in confined fluidic media, are expected to benefit [145]. Applications related to biomedical devices and synthetic biology are also anticipated, where controlled transport is relevant in the case of biological samples for cell culture, diagnostic kits, organic synthesis, drug delivery, and creation of bio-inspired materials or artificial cellular systems [29, 38, 53]. In these technologies, active actuation (*e.g.* external fields) might offer additional leverage in enhancing control and efficiency of these processes.

Applications relating to heat transfer and thermal management can greatly benefit from the possibility of droplet control. Here, anticipated applications are related to enhanced condensation heat transfer, anti-icing, fog harvesting, and advanced cooling systems [4, 83, 110, 113, 154, 167]. Self-cleaning and antifouling surfaces are expected to benefit from a more precise control of droplet motion, where surfaces can spontaneously remove contaminants, which is crucial for optical sensing and materials operating in extreme environments [154]. To this end, a range of functional surfaces can be envisioned, which can also adapt and respond to external stimuli or internal gradients inspired by biological processes that can lead to 'smart' windows, liquid diodes, and logic gates [5, 53].

The research area of directed-fluid motion of droplets and cells as discussed in the context of this review is rapidly evolving. To accelerate innovation in this direction, different approaches can be pursued. One possibility is the integration of multi-modal gradients. This refers to the combination of multiple transport mechanisms based on different gradients (*e.g.* chemical, topographic, electrical, thermal, *etc.*) to achieve a finer control and efficiency in terms of distance covered and velocity of the droplets. More sophisticated functionalities may include the possibility of switching between the different processes and adjusting the relative strengths of the different factors to acquire a finer control over their synergistic effect. Additional control can be achieved by combining active and passive transport processes, which can further broaden the effects that could be used in the various applications, thus widening the area of applications of gradient-driven motion.

Beyond static gradients, actively reconfigurable surfaces, where gradients in properties along the substrate can be created, removed, or varied in realtime, offer immense potential for adaptive and responsive microfluidic devices. Possibilities for realising such scenarios include photo-responsive surfaces, magnetically actuated surfaces, and soft deformable substrates [62, 133, 137]. To enable applications, further understanding of the fundamental forces acting at complex interfaces and their imbalance is required, such as investigations of the fundamental physics of these processes at solid–liquid, liquid–liquid, and liquid–vapour interfaces, for example, in the case of soft or multi-layered substrates. To this end, contact line dynamics, wetting ridge formation, and the interplay between surface tension, elasticity, and flow fields require detailed investigations that can further enhance our understanding of the related phenomena [131, 158, 168].

In the context of biologically related processes, such as durotaxis and other mechano- and chemico-sensing phenomena, progress shall be made in translating the *in vitro* findings into *in vivo* demonstrations [88]. This highly relies on developing sophisticated tools and models that can recapitulate the complex, dynamic, and heterogeneous mechanical environments found in living organisms [28].

Another aspect for applications at macroscales is linked to the ability of scaling up current designs that have been used in laboratory environments to the pilot and application level, when this is necessary. Translating laboratory-scale demonstrations into practical applications requires scalable and cost-effective fabrication methods to create surfaces with tailored gradients and textures [112, 113]. The user-end applications shall also be sustainable and energy efficient. In this regard, passive transport mechanisms and self-powered systems (*e.g.* triboelectric generators [56]) seem to offer certain advantages over active processes that require a continuous energy supply to maintain the droplet motion, which is important for developing sustainable and low-energy microfluidics and environmental technologies. Moreover, autonomous intelligent systems may rely on gradient substrates. Here, the ultimate goal is to intelligently drive droplets or cells along complex trajectories. Integrating AI methods could offer further possibilities for predictive control and adaptation towards defect-free applications of gradient-driven motion, as many of the related processes rely on a number of different parameters that may act synergistically. In all the above applications, the main ideas discussed in the context of Figure 1 remain central, namely the imbalance of forces along the droplet–substrate interface will drive the droplet motion in the case of passive transport, and can compete with others, such as viscous forces. These imbalances should be tuned in a way to increase the efficiency of the motion in terms of droplet velocity, acceleration, distance covered by the droplet and stability of the motion.

5. Concluding remarks

Research on directed droplet-motion onto substrates with gradient properties has attracted much interest over the last two decades due to the continuous miniaturisation of technologies, where tiny effects taking place at the interface between liquids and substrates can be exploited in applications, such as digital microfluidics, and medicinal diagnostics. The research in the area of gradient substrates also offers exciting opportunities for fundamental scientific discoveries and the development of transformative technologies across diverse fields. It can further be combined with various other processes that rely on changes of properties stimulated by different driving forces, such as aerotaxis (stimulation by oxygen), anemotaxis (by wind), barotaxis (by pressure), chemotaxis (by chemicals), durotaxis (by stiffness), electrotaxis of galvanotaxis (by electric current), hydrotaxis (by moisture), magnetotaxis (by magnetic field), phototaxis (by light), rheotaxis (by fluid flow), thermotaxis (by temperature changes), thigmotaxis (by physical contact). These offer further possibilities for droplet actuation and control, with applications extending to the macroscopic scales. It is therefore anticipated that the research area of directed fluid flow will further be enriched by new possibilities for realizing such motion and its control, as well as continue to draw further inspiration from observed natural phenomena.

Acknowledgements

This research was supported by the National Science Centre, Poland, under grant No. 2019/35/B/ST3/03426. A. M. acknowledges support by COST (European Cooperation in Science and Technology [See http://www.cost.eu and https://www.fni.bg] and its Bulgarian partner FNI/MON under KOST-11).

vapor film below the leidenfrost temperature, ACS Nano 12 (2018) 11995–12003. doi:10.1021/acsnano.8b04039.

[168] X. Dai, B. B. Stogin, S. Yang, T. S. Wong, Slippery wenzel state, ACS Nano 9 (2015) 9260–9267. doi:10.1021/acsnano.5b04151.
48